\documentclass[aps,preprint,showpacs,nofootinbib,12pt]{revtex4}
\usepackage[dvips]{color}
\usepackage{CJK}
\usepackage{indentfirst}
\usepackage{bm}
\usepackage{ctex}
\usepackage{amsmath}
\usepackage{amssymb}
\usepackage{amstext}
\usepackage{amsfonts}
\usepackage{graphicx,epsfig}
\usepackage{slashbox}
\usepackage{indentfirst}

\begin{document}
\title{Effects of nuclear deformation on the form factor for direct dark matter detection }
\author{Ya-Zheng Chen, Jun-Mou Chen, Yan-An Luo, Hong Shen
and Xue-Qian Li       }

\affiliation{School of Physics
 NanKai University, Tianjin 300071, China }

\begin{abstract}
\noindent For direct dark matter detections, to extract useful
information about the fundamental interaction from data, it is
crucial to properly determine the nuclear form factor. The form
factor for spin-independent cross section of collisions between dark
matter particle and nucleus is thoroughly studied by many authors.
When the analysis was carried out, the nuclei are always supposed to
be spherically symmetric. In this work, we investigate the effects
of deformation of nuclei from a spherical shape to an elliptical
shape on the form factor. Our results indicate that as long as the
ellipticity is not too large, such effects cannot cause any
substantial effects, especially as the nuclei are randomly
orientated in a room temperature circumstance one can completely
neglect them.

\end{abstract}
\pacs{21.60-n, 24.10.Cn,  95.35+d }

\maketitle
\section{Introduction}

With serious astronomical observation of several decades, existence
of dark matter is no longer doubtful. On another aspect,  we
definitely know that in the zoo of the standard model (SM) we do not
have any candidates for dark matter (DM). Question is what the Dark
Matter particles are. There are many models proposed in
literature~\cite{He:2011zzi,Briscese:2011zz,Cheung:2011zza,SL2010,NSA},
but unless they are caught by our detectors in the terrestrial
laboratories or
satellites~\cite{Morselli:2011zz,Profumo:2009zz,CMS}, one still
cannot surely identify them. Much efforts have been made to discover
the dark matter flux from outer space.

Comparing with the spin-dependent cross section, the
spin-independent cross section of dark matter particle with nucleus
is much larger due to the $A^2$ enhancement where $A$ is the atomic
mass number of the nucleus as the detection
material~\cite{Jungman:1995,witten,Ressell:1993qm,Ressell:1997kx,Engel:1989ix,Engel:1991wq,Drees:2008qg}.
Even so, the cross sections of elastic scattering between DM
particles and nuclei are still small, the present experiments have
already reached $10^{-44}$ cm$^2$. Because of the advantage of the
spin-independent scattering whose cross sections are larger and the
theoretical treatment is relatively simpler than that for
spin-dependent processes, nowadays, the priority of research is
given to the study on spin-independent elastic DM-nucleus reactions.

Since the kinetic energy of the DM particle is rather low at order
of a few tens of keV, the impact of DM particle on nucleus is almost
impossible to cause inelastic processes, thus all observational
signals are related to the recoil of nucleus after the collision.
Namely, even the collision occurs between DM and quarks (seldom
gluons in the case), all the absorbed energy is passed to the
nucleus to make it to move as a whole object, the recoil which may
induce thermal, electric and light signals. For the spin-independent
cross section, the particle-physics and nuclear-physics
contributions can be separated, namely the nuclear effects can be
factored out and included in a form factor $F({\bf q})$. For
spherically symmetric nuclei, $F({\bf q})$ only depends on $|{\bf
q}|$. The spherical symmetry means that the nuclei are of full-shell
or close to full-shell structures, but for most of the nuclei which
are taken as the detection materials the shells are not completely
filled out. A careful study on form factors for the not-full-shell
structure nuclei would be helpful for extracting information about
fundamental interactions from the data. For that case, $F({\bf q})$
is not only a function of $|{\bf q}|$, but also $\cos\theta$ while
the azimuthal symmetry is assumed. We will write it as
$F(q,\cos\theta)$ where $q\equiv |{\bf q}|$.

Nucleus is a complex many-body system, therefore extraction from
data requires a thorough analysis on the nuclear structure. The form
factors for spherical nuclei have been carefully studied by many
authors and the results can be applied to analysis of data. In this
work, we are going to investigate the effects of deformation of
nuclei on the form factor, namely we will derive the form factors
corresponding to the deformed nuclei with relatively smaller
ellipticity.

We employ several models to calculate the form factors
$F(q,\cos\theta)$ for nuclei with small ellipticity. We will take
$Xe$ and $Ge$ which are commonly adopted as the detection materials
as examples to illustrate the effects of deformation.

The paper is organized as follows, after this introduction, we
present the expressions of the form factors derived from different
models for nuclear density, and then present our numerical results
via several figures. The last section is devoted to our conclusion
and discussion.

\section{The form factor related to deformed nucleus}

Obviously it is reasonable to assume that the nucleus with a larger
$A$ may be only quadruply deformed, namely  it is deformed from
spherical form to ellipsoidal. In the spherical coordinates, the
nuclear density of a nucleus with an elliptical form should be
$\rho(r,\theta)$ which is a function of both radius $r$ and polar
angle $\theta$ and the corresponding form factor should be written
as follows, it needs to be noted that we would set
$\varphi_1=\varphi_2$ in practical calculation for simplifying the
integration.

\begin{eqnarray}
F(q,\theta_2)&=&\frac{1}{M}\int \rho(r,\theta_1)e^{i\vec q \cdot
\vec r} d^3r\nonumber\\
&=&\frac{1}{M}\int_{0}^{\pi}\sin\theta_1 d\theta_1
\int_{0}^{2\pi}d\varphi_1 \int_{0}^{\infty} \rho(r,\theta_1)
e^{iqr(\sin\theta_1 \sin\theta_2
\cos(\varphi_1-\varphi_2)+\cos\theta_1 \cos\theta_2)} r^2 dr .
\end{eqnarray}

Even though this work is aiming to find the effects of deformation
of nuclei on the form factor, the deviation from the spherical form
for the nuclei under investigation is not severe, therefore, we can
always start from a spherical form and then make reasonable
modifications or extension.

\subsection{Extension of the Two-Parameter Fermi Distribution(E2PF) }
A number of models have been proposed
\cite{Duda:2006uk,Chen:2011xp} to describe the nuclear charge
density or mass density. Among them the two-Parameter Fermi
distribution(2PF) is one of the simple models. For a spherical form
the density is written as
\begin{equation}
\rho(r)=\frac{\rho_0}{1+\exp(\frac{r-c}{z})} \label{2PF},
\end{equation}
where $\rho_0 $ is equal to $2\rho(r)$ at $r=c$, and $z$ is the
diffusivity of the surface. It would be convenient for later use to
derive the mean square root radius $\bar R$ for the spherical 2PF
model as
\begin{equation}
\bar R^{2PF}=\sqrt{\frac {4\pi\int_0^\infty r^4\rho(r)dr}{4\pi
\int_0^{\infty}r^2 \rho(r)dr}}=\sqrt{
\frac{3}{5}c^2+\frac{7}{5}\pi^2 z^2}.
\end{equation}

For an elliptical nucleus with an axial symmetry, the nuclear
density in the two-parameter Fermi distribution (E2PF) model should
be extended as
\cite{Greiner,Ismail,Cooper,Xu,walecka,Hahn,Hohr,Ismail:2008zz}:
\begin{equation}
\rho(r,\theta)=\frac{\rho_{0}'}{1+\exp \left(
\frac{r-c(\theta)}{z}\right) }, \label{E2PF}
\end{equation}\label{rho0}
where
\begin{equation}\label{RT}
c(\theta )=c_{0} (1+\beta _{2}Y_{20}(\theta)).
\end{equation}
The parameter $\beta_2$ which corresponds to the ellipticity of the
nucleus characterizing its deformation from a spherical form, is a
small quantity for the nucleus which we concern. For  the priori
assumption of small deformation, we only keep the multiple terms up
to $Y_{20}$ \cite{Ismail:2008zz}.

The parameter $c_0=1.1A^{1/3}$ and  $\rho_0'$ can be obtained by the
normalization condition i.e. requiring the integration over whole
coordinate space to be equal to the nuclear mass number (or total
charge Ze) which is priori set for various nuclei. $\beta_2$ can be
obtained from the data book \cite{data}, and it is  $-$0.113 and
$-$0.224 for $^{131}Xe$ and $^{73}Ge$ respectively. $z$ denotes the
surface diffuseness. Here we choose the normalization as follows:
\begin{equation}
\int \rho(r,\theta)d^3 r=M.
\end{equation}

\begin{figure}[htbp]
\begin{center}
\includegraphics[width=12cm ]{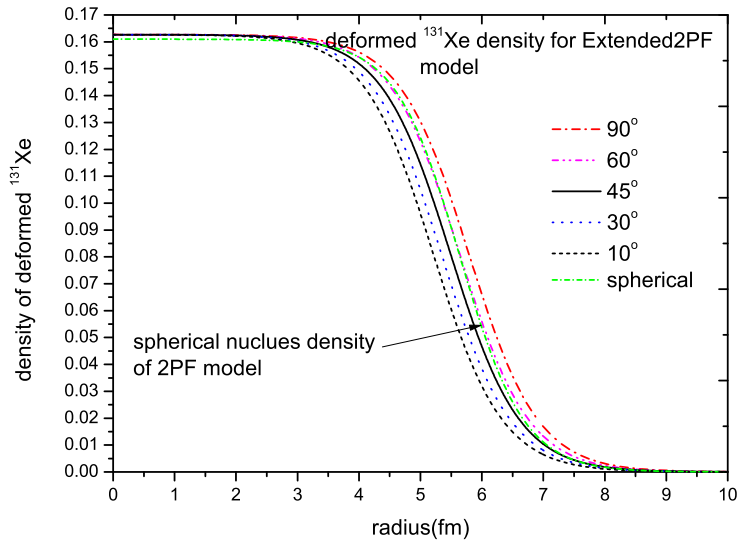}
\caption{Nuclear density of $^{131}$Xe  for the Extended 2PF model. It
shows the change of density with increasing angle from 10$^\circ$ to
90$^\circ$, the Short Dash Dot line(green) corresponds to the case of
the spherical 2PF model  }
\label{E2PFden}
\end{center}
\end{figure}
\begin{figure}[htbp]
\begin{center}
\includegraphics[width=12cm ]{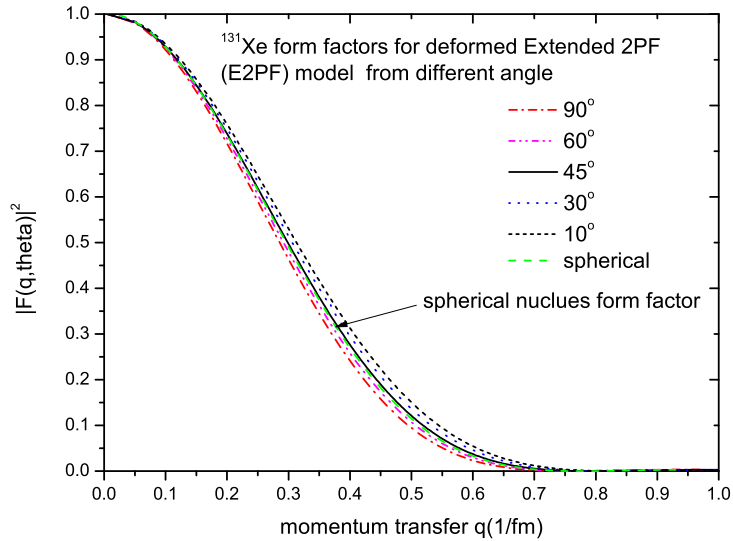}
\caption{$^{131}$Xe form factors for deformed nucleus Extended
2PF(E2PF)model from different directions: 10$^o$, 30$^o$, 45$^o$,
60$^o$, 90$^o$ }
 \label{E2PFF}
\end{center}
\end{figure}
In Fig.($\ref{E2PFden}$) we show the density distribution for
$^{131}$Xe in the E2PF model. It is observed that from the center of
the nucleus to about three fermis, the density remains unchanged in
all directions. Then the angular distribution of the density begins
to be apart for different angles beyond three fermis. The short dash
dot(green) line is the 2PF density model when the nucleus is assumed
to be spherical. Fig.($\ref{E2PFF}$) is the corresponding form
factors, which are calculated by taking a Fourier transformation to
the deformed nuclear density in the configuration space.
\subsection{Extension of the Folding model(EF)}

There is another commonly adopted model which is rather simple, i.e.
the nucleons are postulated to be uniformly distributed in a sphere
with a certain boundary radius. For an axially symmetric ellipsoidal
shape, one should extend the density for a spherical form. The
surface equation of an ellipsoid is
\begin{eqnarray}\label{a1}
\frac{x^2}{a^2}+\frac{y^2}{a^2}+\frac{z^2}{b^2}=1,
\end{eqnarray}%
or in the spherical coordinate system, it is written as
\begin{eqnarray}
R(\theta)=\sqrt{\frac{a^2 b^2}{(a^2 - b^2)\cos^2\theta +b^2}}.
\label{rt}
\end{eqnarray}
In an approximation, if we only keep the multiple terms to
quadrupole, we can re-parametrize the surface equation to a more
convenient one
\begin{equation}\label{radius}
R(\theta)=R_0(1+\beta_2
Y_{20}).
\end{equation}

Extending the Folding model,  we  set the nuclear density to be
uniform inside the ellipsoid with radius $R(\theta)$
\begin{equation}\label{rho_0}
\rho_0(r,\theta)=\frac{3M}{4\pi a^2 b}\Theta(r-R(\theta)),
\end{equation}
where $\Theta$ is the step function. Following the
literature~\cite{Helm}, we introduce a smearing function $\rho_1$ to
take care of  the soft edge effect of the nucleus:
\begin{equation}\label{rho1}
 \rho_1(r)=\frac{1}{(2 \pi s^2)^{3/2}}\exp(\frac{-r^2}{2s^2}),
\end{equation}
then one should convolve $\rho_0$ and $\rho_1$ to get the nuclear
density
\begin{eqnarray}
\rho(r,\theta)&=& \int \rho_{0}(\vec{r'}) \rho_1(\vec{r}-\vec{r'}) d^3 r' \nonumber\\
              &=&\int \rho_{0}(\vec{r'}) \rho_1(\vec{r}-\vec{r'})
              r'^2 dr' \sin{\theta'} d\theta' d\varphi'  \nonumber\\
              &=& \frac{ 1}{(2\pi s^2)^{3/2}} \int_0
^{2\pi} d \varphi'\int_{0}^{ \pi} \sin{\theta'} d\theta'
\int_{0}^{\infty} \frac{3M}{4\pi a^2 b} \Theta(r'-R(\theta')) \times
\nonumber\\
&&\exp(\frac{-(r^2 +r'^2- 2rr'(\sin\theta \sin\theta'
\cos(\varphi-\varphi')+\cos\theta' \cos \theta))}{2s^2}) r'^2 dr'.
\end{eqnarray}
The  semi-axes $a$  and $b$ are set as
\begin{eqnarray}
a&=&R(\theta=\frac{\pi}{2})=R_0(1+\beta_2
Y_{20}(\frac{\pi}{2}))=R_0(1-\sqrt{\frac{5}{16\pi}}\beta_2)\nonumber\\
b&=&R(\theta=0)=R_0 (1+\beta_2
Y_{20}(0))=R_0(1+2\sqrt{\frac{5}{16\pi}}\beta_2).
\end{eqnarray}

In the extended Folding (EF) model, we can also calculate the mean
square root radius $\bar R^{EF}$. For a spherical nucleus, one may
equate the mean square root radius obtained in the 2PF and Fold
models, thus he acquires the spherical radius $R_0$ for the Fold
model \cite{Duda:2006uk,JDL}.
\begin{equation}\label{R0}
R_0=\sqrt{c^2+\frac{7}{3}\pi^2 a^2 -5s^2},\quad c \simeq
(1.23A^{1/3}-0.6 ) \mbox{fm},  \quad s=0.9   \mbox{fm}, a=0.52
\mbox{fm}.
\end{equation}

As aforementioned, the deformation makes the shape of the nucleus
slightly deviate from a spherical form, we can still use the above
relation achieved for spherical nuclei and set $R_0$ to be the
parameter in Eq.(\ref{radius}).

A Fourier transformation would bring the nuclear density to the
expected form factor $F(q,\theta)$. It is noted that now the form
factor is also direction-dependent.
\begin{eqnarray}
F(q)&=&\int \rho_0({\bf r}') \rho_1({\bf r}- {\bf r}' )d^3r' e^{i {\bf q} \cdot {\bf r}}d^3r \nonumber\\
&=& \int \rho_0({\bf r}')d^3r' \int \rho_1({\bf r}- {\bf r}' )d^3r' e^{i {\bf q} \cdot {\bf r}}d^3r\nonumber\\
&=&\int_{0}^{\infty} \rho_0({\bf r}')d^3 r' \int_{0}^{\infty} \rho_1({\bf u}) e^{i{\bf q} \cdot {\bf r}}d^3u \
\quad  (\mbox{setting} \quad {\bf r}-{\bf r}'={\bf u}) \nonumber\\
&=& \int_{0}^{\infty} \rho_0({\bf r}')d^3 r'\int_{0}^{\infty}\rho_1({\bf u}) e^{i {\bf q} \cdot({\bf u}+{\bf r}')} d^3u \nonumber\\
&=&\int_{0}^{\infty} \rho_0({\bf r}')e^{i{\bf q} \cdot {\bf r}'}d^3r'\int_{0}^{\infty} \rho_1({\bf u})e^{i {\bf q} \cdot {\bf u}} d^3u \nonumber\\
&=& F_0(q)F_1(q),
\end{eqnarray}
and
\begin{eqnarray}
F_1(q)&=&\int \rho_1(r)e^{i {\bf q} \cdot {\bf r}}d^3r \nonumber\\
&=&e^{-{\bf q}^2/2}.
\end{eqnarray}
We take a trick to make the integration easier as ${\bf r}$ is
described in the cylindrical coordinate while ${\bf q}$ is described
in the spherical coordinate as:
\begin{displaymath}
\left\{ \begin{array}{l}
     x=t\cos \varphi_1\\
     y=t \sin\varphi_1 \\
     z=z
   \end{array}\right \} \quad
\left\{ \begin{array}{l}
     q_x=q \sin \theta_2 \cos \varphi_2\\
    q_y=q \sin \theta_2 \sin \varphi_2 \\
     q_z=q \cos \theta_2
   \end{array}\right \},
\end{displaymath}
then
$${\bf q}\cdot{\bf r}=t \cos\phi_1 q\sin\theta_2\cos\phi_2+t \sin\phi_1
q\sin\theta_2\sin\phi_2+z q\cos\theta_2,$$ where
$t=\sqrt{x^2+y^2}=\sqrt{r^2-z^2}$.

Thus we obtain
\begin{eqnarray}
F_0(q,\theta_2)&=&\int \rho_0(r,\theta)e^{i {\bf q} \cdot {\bf r}}d^3r \nonumber\\
&=& \int \frac{3M}{4\pi a^2 b} e^{i(x q_x+y q_y+z q_z)} t dt d \varphi_1 dz \nonumber \\
&=& \frac{3M}{4\pi a^2 b} \int e^{i (q \sin \theta_2 \cos \varphi_2
t \cos \varphi_1 + q \sin
\theta_2 \sin \varphi_2 t \sin \varphi_1 + q \cos \theta_2 z)} t dt d\varphi_1 dz \nonumber\\
&=& \frac{3M}{4\pi a^2 b} \int_{0}^{a} e^{iq(t \sin \theta_2 \cos
(\varphi_2-\varphi_1)+z \cos \theta_2)}t dt \int
_{-\sqrt{(1-\frac{t^2}{a^2})b^2}}^{\sqrt{(1-\frac{t^2}{a^2})b^2}}dz
\int _{0}^{2\pi}d\varphi_1.
\end{eqnarray}
The parameters $a$ and $b$ are the semi-axes defined above. Thus the
form factor in the EF model can be written as:
\begin{equation}
F(q,\theta_2)=F_0(q,\theta_2)F_1(q).
\end{equation}

\begin{figure}[htbp]
\begin{center}
\includegraphics[width=12cm ]{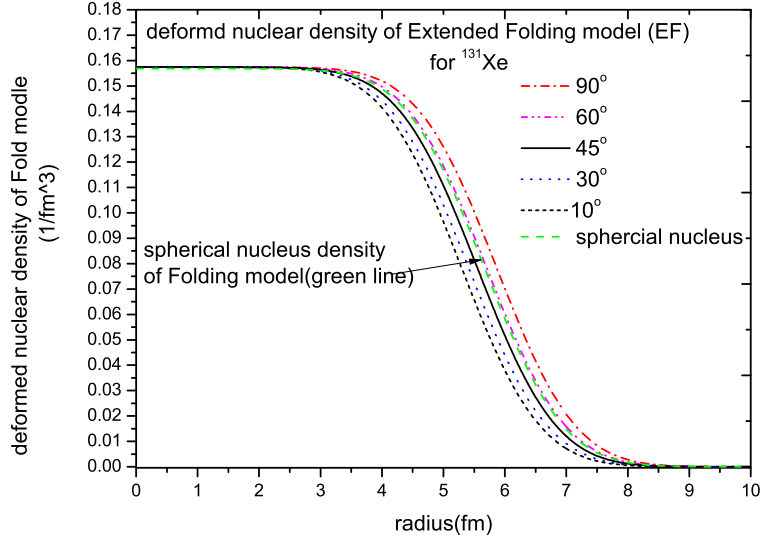}
\caption{$^{131}$Xe form factors for deformed nucleus  of Extended
2PF(E2PF)model from different directions: 10$^o$, 30$^o$, 45$^o$,
60$^o$, 90$^o$ } \label{folddensity}
\end{center}
\end{figure}

In Figure $\ref{folddensity}$, the $^{131}$Xe density distribution
determined by the EF model is shown, while the corresponding form
factors are given in Fig.4. The short dash dot(green) line
corresponds to the spherical form of the nucleus. For the spherical
nucleus, it has already been known that the form factor obtained
with the 2PF model is very close to that determined by the Folding
model\cite{Chen:2011xp}. Thus we will also make a comparison between
the E2PF and EF form factors at the end of the paper. Then we will
present the third model to do the same job in the following section.
\begin{figure}[htbp]
\begin{center}
\includegraphics[width=12cm ]{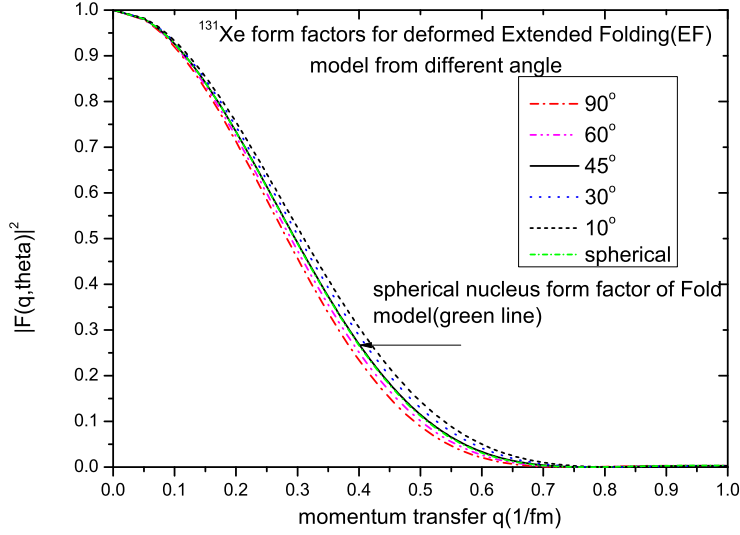}
\caption{$^{131}$Xe form factors for deformed nucleus determined in the EF
model for different directions: 10$^o$, 30$^o$, 45$^o$,
60$^o$, 90$^o$ } \label{foldFF}
\end{center}
\end{figure}

\subsection{The Nilsson Mean Field(NMF) }

Above, we use two simplified models (E2PF and EF) to derive the form
factors for deformed nuclei. The advantage is that the models are
simple and we can obtain analytical solution which is convenient for
illustrating the characteristics of the form factors, but might be
too simplified. Now we turn to use a more realistic model.

In this subsection,  the form factors for  deformed nuclei are
obtained in the Nilsson modified oscillator model, then using
$^{131}$Xe as an example, we present the results in some figures.

Below, let us briefly review the model and show how we apply it to
study the concerned form factor.


In the Hamiltonian of the Nilsson model the potential for an axially
symmetric harmonic oscillator can be written as
\cite{Nilsson1969,Nilsson1958,Nilsson1955}
\begin{equation}
H={-\hbar^2\over 2M}\nabla^2 +\frac{1}{2} M[\omega_x^2(x^2+y^2)+\omega_z^2z^2]-C{\bf s}\cdot {\bf l} -D{\bf l}^2,
\end{equation}
where $C{\bf s}\cdot {\bf l}$ is the spin-orbit coupling, and $D{\bf
l}^2$ flattens the bottom of the potential.

A deformation parameter $\delta$ is introduced to reflect the axial
symmetry for the deformed nuclei as
\begin{eqnarray}
\omega_x^2 &=& \omega_y^2=\omega_0^2(\delta)(1+\frac{2}{3}\delta),\\
\omega_z^2 &=& w_0^2(\delta)(1-\frac{4}{3}\delta).
\end{eqnarray}

The equipotential surface encloses a constant volume if
\begin{equation}
\omega_x\omega_y\omega_z=const.
\end{equation}
Then we have
\begin{equation}
\omega_0[1-\frac{4}{3}\epsilon_2^2-\frac{16}{27}\epsilon_2^3]^{1/6}=\omega_{00}.
\end{equation}

The Hamiltonian  can be decomposed into three pieces as
\begin{eqnarray}
H &=& H_{sp}+H_{\epsilon_2}-C{\bf s}\cdot{\bf l}-D{\bf l}^2,\\
H_{sp}&=&\frac{p^2}{2m}+\frac{1}{2}m\omega_0^2r^2,\\
H_{\epsilon_2}&=&-m\omega_0^2r^2\frac{2}{3}\epsilon_2P_2(cos\theta).
\end{eqnarray}

It would be convenient to use dimensionless coordinates and
parameters which are defined as
\begin{eqnarray}
\rho&=&\sqrt{\frac{m\omega_0r}{\hbar}}r,  \\
C &=& 2\kappa\hbar\omega_{00}, \\
D&=&\frac{1}{2}C\mu=\kappa\hbar\omega_{00}\mu.
\end{eqnarray}

Then the Nilsson Hamiltonian can be further written as
\begin{eqnarray}
H&=&\hbar\omega_0(H_0-\frac{2}{3}\epsilon_2P_2)-\kappa\hbar\omega_{00}[2{\bf s}\cdot {\bf l}+\mu({\bf l}^2-<{\bf l}^2>_N)],\\
H_0&=&\frac{1}{2}(-\nabla_\rho^2+\rho^2),
\end{eqnarray}
where $<{\bf l}^2>=N(N+3)/2$ is an average over all states within
the $N-$th shell, and $\hbar\omega_{00}\approx 41A^{-1/3}$ MeV.

If the octupole and hexadecupole deformations are considered, the
Hamiltonian would become more complicated as
\begin{equation}
H=\hbar\omega_0(H_0+\rho^2(-2/3\epsilon_2P_2+\epsilon_3P_3+\epsilon_4P4))-\kappa\hbar\omega_{00}[2{\bf
l}\cdot{\bf s}+\mu ({\bf l}^2-<{\bf l}^2>_N)]
\end{equation}
and it is the Hamiltonian we are going to use in the later part of
this paper.

The Nilsson wavefunction is constructed with the spherical harmonic
oscillator basis $|Nlj\Omega>$,
\begin{equation}\label{TT}
\Psi_i=\sum_{\substack{\alpha}}\omega_{\alpha}c_{{\alpha}}^{\dag}|0>,
\end{equation}
where $\omega_{\alpha}$ is a coefficient, $\alpha$ refers to a set
of quantum numbers $(njl\Omega)$ of the harmonic-oscillators.

The nuclear density thus is expressed as
\begin{equation}
\rho=\sum_{\substack{i=1}}(\Psi_{i}^\dag(\pi) \Psi_{i}(\pi)
+\Psi_{\bar i}^\dag(\pi)
\Psi_{\bar{i}}(\pi))+\sum_{\substack{i=1}}(\Psi_{i}^\dag(\nu)
\Psi_{i}(\nu) +\Psi_{\bar i}^\dag(\nu) \Psi_{\bar{i}}(\nu)),
\end{equation}
where $\bar{i}$ represents  the time-reversed states.

In this paper, the major shells under consideration are from 0 to 9
for proton and neutron, respectively. The quadrupole, octupole and
hexadecapole deformation parameters are determined by
experiments\cite{nndc}, which are  -0.108,     0 and  0.027,
respectively.

\begin{figure}[t]
\begin{center}
      \includegraphics[angle=0,width=0.48\textwidth]{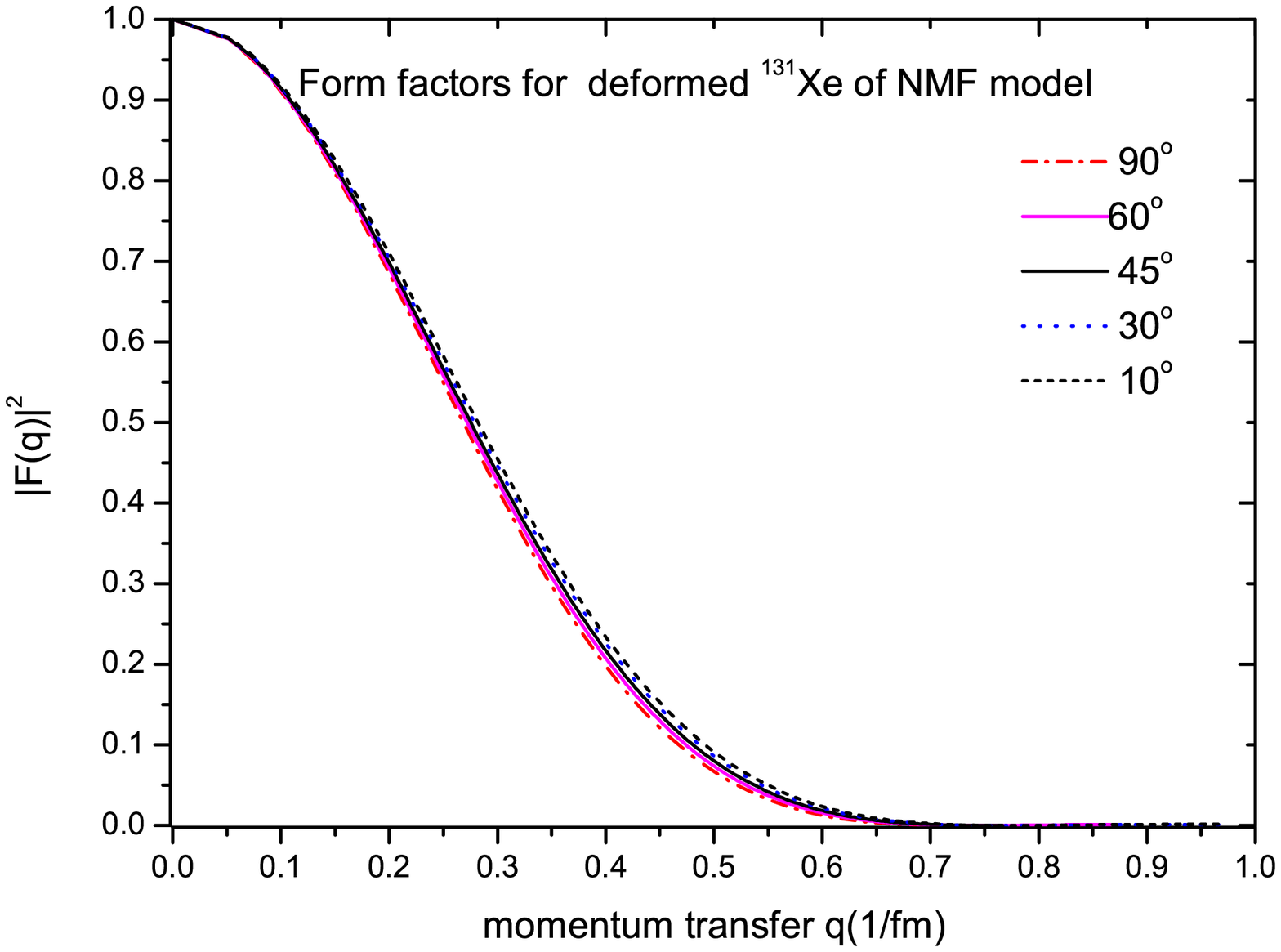}
      \includegraphics[angle=0,width=0.48\textwidth]{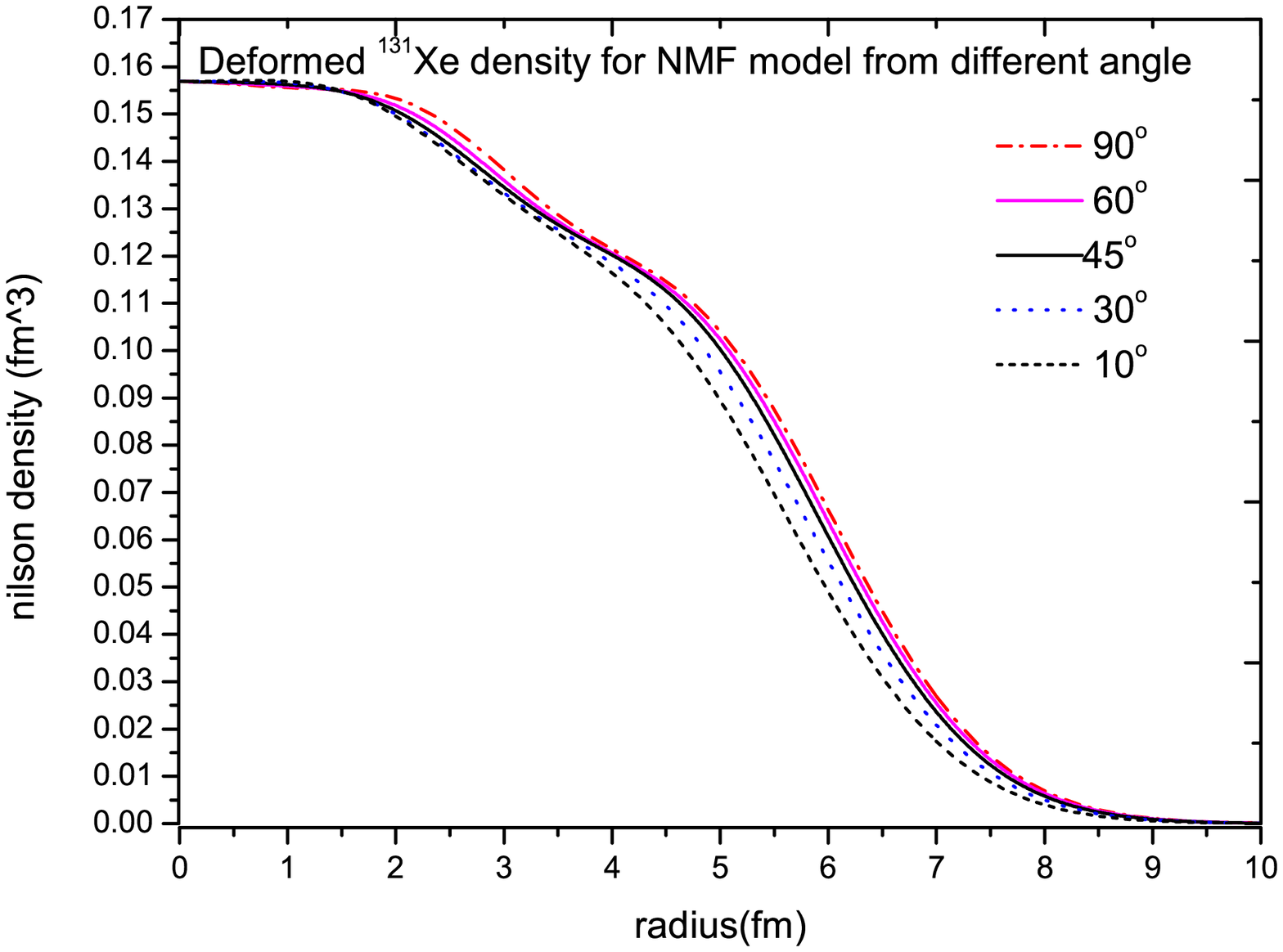}
\caption{ The right panel shows the dependence of the $^{131}$Xe density
on the directions from 10$^{o}$ to 90$^{o}$ obtained in the
Nilsson Mean Field model, the left panel is the form factor}
\label{NIlssonXe}
\end{center}
\end{figure}
\begin{figure}[h]
\begin{center}
   \includegraphics[angle=0,width=0.48\textwidth]{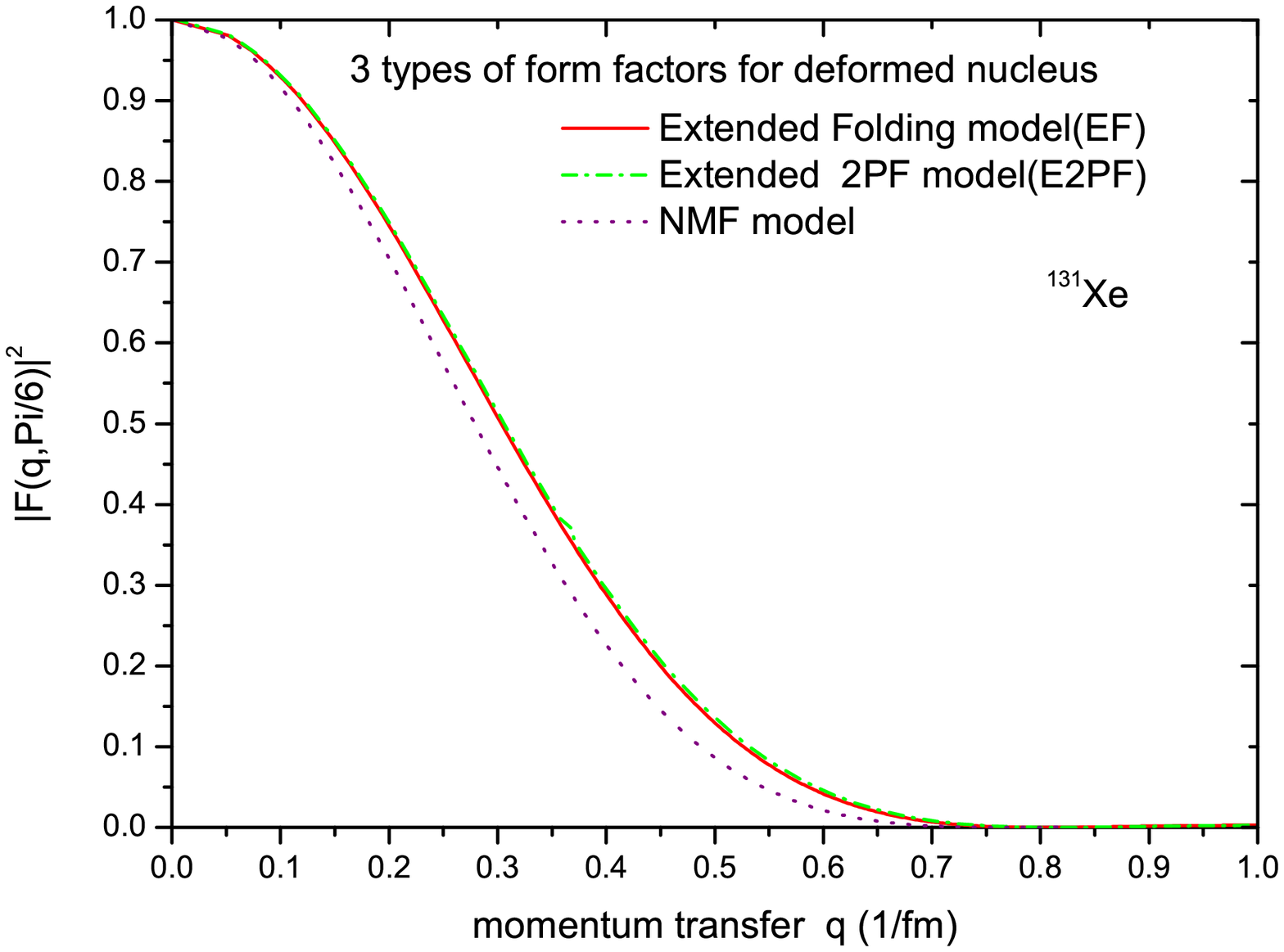}
   \includegraphics[angle=0,width=0.48\textwidth]{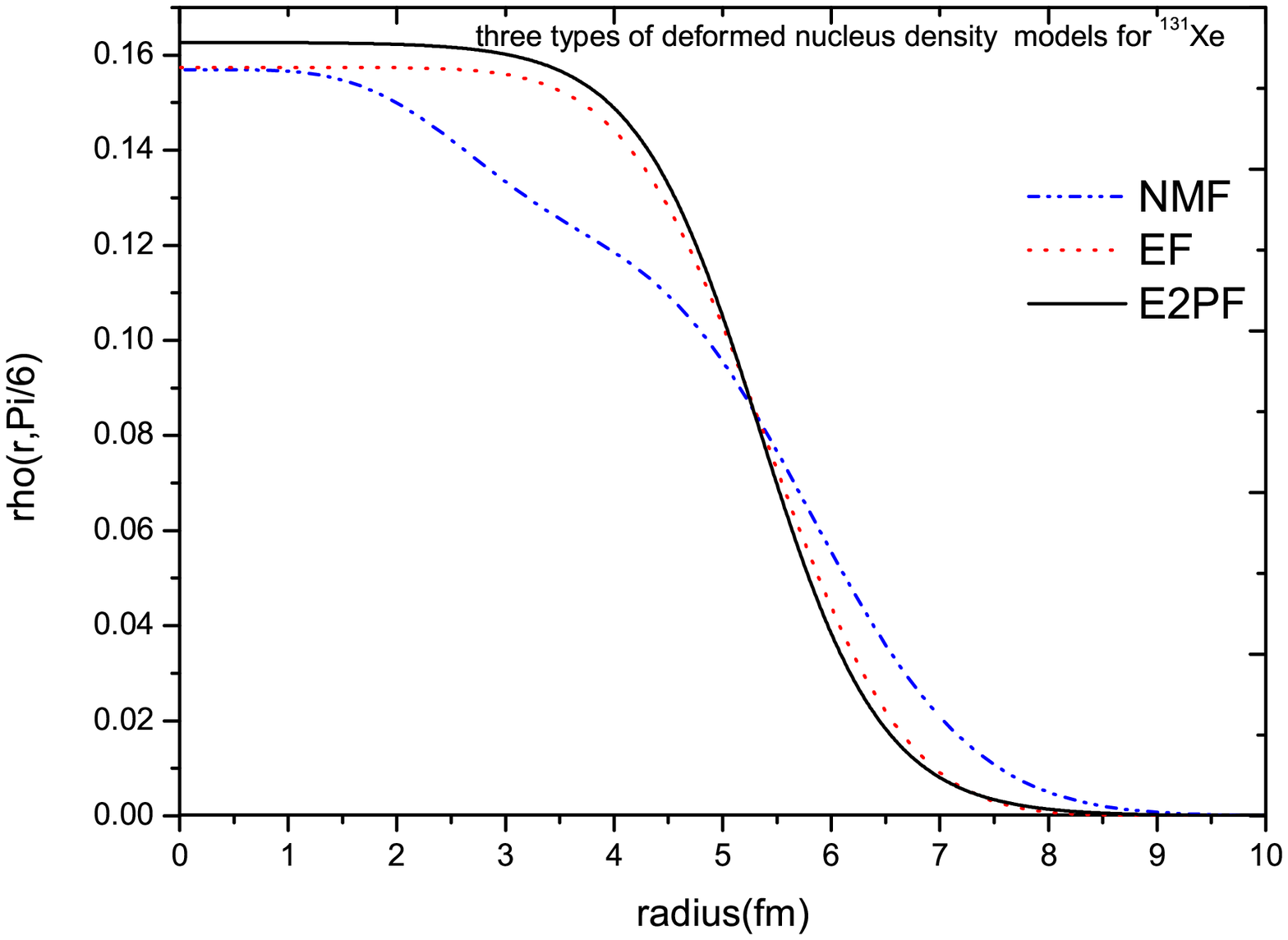}
  \caption{The form factors F(q,$\pi/6$) and density $\rho(r,\pi /6)$ obtained in three different models: E2PF,EF,NMF for $^{131}$Xe}
\label{3M}
\end{center}
\end{figure}
\begin{figure}[h]
\begin{center}
    \includegraphics[angle=0,width=0.48\textwidth]{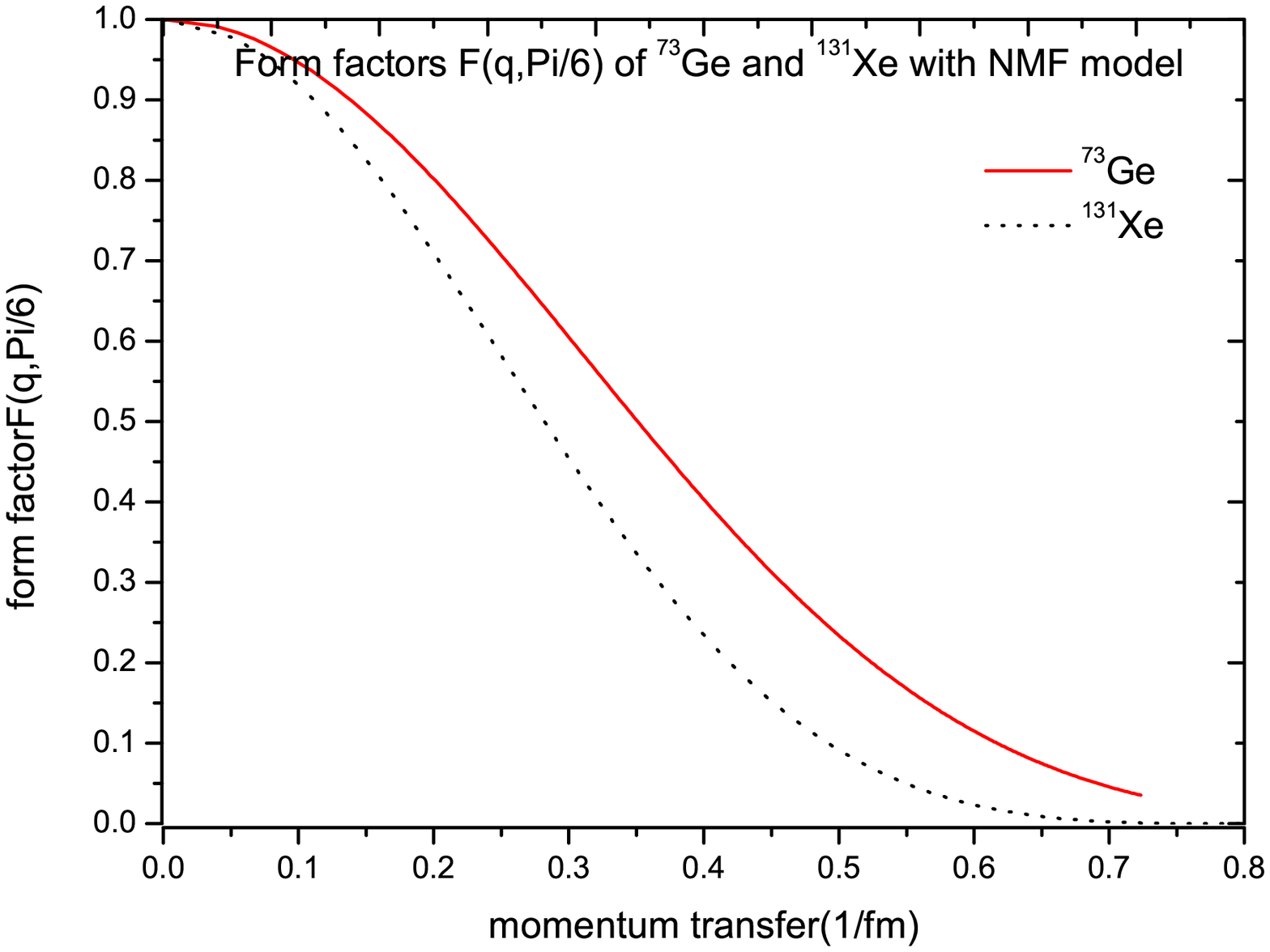}
    \includegraphics[angle=0,width=0.48\textwidth]{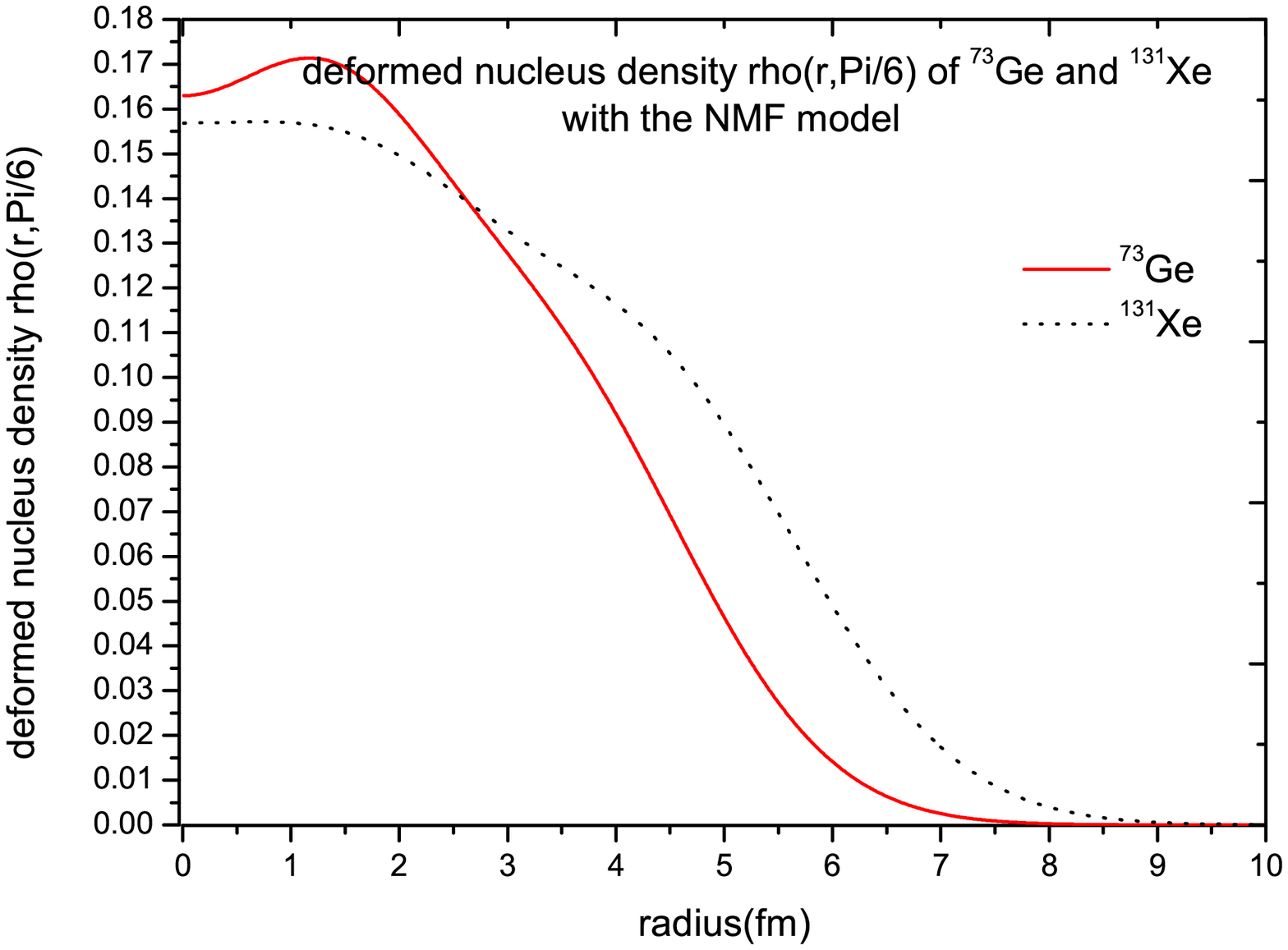}
  \caption{the left graph is the form factors F(q,$\pi/6$) for the $^{73}$Ge and $^{131}$Xe with NMF model, the right graph is  the density distributions $\rho(r,\pi/6)$.  }
\label{GeXe}
\end{center}
\end{figure}

Performing a Fourier transformation on the nuclear density, we
obtain the concerned form factor.  On the right panel of
Fig.($\ref{NIlssonXe}$), the angular dependence of the nuclear
density of $^{131}$Xe  is shown. Fig.($\ref{NIlssonXe}$) plots the
NMF form factors F$(q,\theta)$ with various angles. The left panel
of Fig.($\ref{3M}$) compares the form factors at direction of
$\theta=\pi/6$ obtained with the three models: the EF, E2PF, and
NMF, whereas the right panel is the corresponding densities.
Fig.($\ref{GeXe}$) shows difference of the form factors F$(q,\pi/6)$
for $^{73}$Ge and $^{131}$Xe, as well as their density
distributions.

\section{Summary}

The aim of this work is to study if a small deformation of nuclei
can induce observable effects on the form factors for the direct
detection of dark matter flux. The form factor, no matter the nuclei
are of spherical or deformed shapes, say ellipsoidal, must satisfy
two normalization conditions. First, the nuclear density must be
normalized as
\begin{equation}
\int\rho(r,\theta,\phi)d^3 r=M,
\end{equation}
where $M$ is the total mass of the nucleus. It is independent of the
shape of the nucleus. Secondly, the form factor must satisfy the
condition
\begin{equation}
F(|{\bf q}|=0)=F(0)=1.
\end{equation}
This condition does not depend on polar and azimuthal angles. With
these two conditions, one can adopt different models for the nuclear
density and then carry out a Fourier transformation to convert the
nuclear density from the configuration space into the momentum space
to gain the form factor which corresponds to non-zero momentum
transfer.

In this work, we start with the spherical nuclei and adopt three
models which are commonly employed to study the nuclear effects.
Then we extend them to deformed shapes by including polar angle
dependence in the density while an axial symmetry is assumed for
simplicity. With the three models we obtain the form factor for the
nuclei whose shape slightly deviates from spherical form, namely
their ellipticity is relatively small.

We notice from the figures shown in the text that the form factors
are not very much apart from that for spherical form, indeed the
dependence of the form factor on $|{\bf q}|$ for $\theta=\pi/4$ is
rather close to that for spherical shape.

Especially, if there is not a strong magnetic field to polarize the
nuclei at very low temperature, the nuclei in the detection material
would be randomly oriented, the polar and azimuthal angles would be
averaged and the deformation effects should be eventually smeared
out.

Therefore, our conclusion is that unless one can keep the detector
at very low temperature such as the CDMS detector and apply a strong
magnetic field to it, the effects of deformation of nuclei can be
safely ignored.

\section*{Acknowledgments}
This work is partially supported by the National Natural Science
Foundation of China (NNSFC) under contract No.11075079,
11075080,11075082.

\end{document}